\documentclass{article}
\usepackage{enumitem}
\setlist[itemize]{leftmargin=*}
\usepackage{authblk}
\usepackage{pgfplots}
\usepackage{pgfplotstable}
\usepgfplotslibrary{statistics}
\pgfplotsset{compat=1.18}
\usepackage{etoolbox} 
\usepackage{listings}
\usepackage{xcolor}
\usepackage{adjustbox}
\usetikzlibrary{shapes.geometric, arrows}

\newcommand{\readSingleValueCsv}[2]{%
  \pgfplotstableread[col sep=space, header=false]{#1}\temptable%
  \pgfplotstablegetelem{0}{[index]0}\of{\temptable}%
  \edef\rawVal{\pgfplotsretval}%
  \typeout{File #1 => single average => \rawVal}%
  \global\expandafter\let\csname #2\endcsname=\rawVal%
}

\usepackage[letterpaper,top=2cm,bottom=2cm,left=3cm,right=3cm,marginparwidth=1.75cm]{geometry}

\usepackage[english]{babel}

\usepackage{graphicx}

\usepackage{subcaption}
\usepackage{color}

\usepackage{ifthen}
\usepackage{amsmath,amssymb}

\newboolean{showcomments}
\setboolean{showcomments}{true}
\ifthenelse{\boolean{showcomments}}
{ \newcommand{\mynote}[2]{
     \fbox{\bfseries\sffamily\scriptsize#1}
       {\small$\blacktriangleright$\textsf{\textcolor{red}{{\em #2}\bf }}$\blacktriangleleft$}}}
       { \newcommand{\mynote}[2]{}}

\usepackage{longtable}

\usepackage{amsmath}
\usepackage{graphicx}
\usepackage{hhline}
\usepackage{booktabs}
\usepackage[colorlinks=true, allcolors=blue, breaklinks]{hyperref}
\newcommand{\Z}{\mathbb{Z}}

\title{Privacy Enhanced QKD Networks:\\ Zero Trust Relay Architecture based on Homomorphic Encryption}
\author[1]{Aitor Brazaola-Vicario}
\author[1]{Oscar Lage}
\author[1]{Julen Bernabé-Rodríguez}
\author[2,3]{Eduardo Jacob}
\author[2,3]{Jasone Astorga}
\affil[1]{TECNALIA, Basque Research and Technology Alliance (BRTA), Parque Científico y Tecnológico de Bizkaia, Astondo Bidea, Edificio 700. E-48160 Derio (Bizkaia), Spain}
\affil[2]{Dept. of Communications Engineering, University of the Basque Country (UPV/EHU), Bilbao, Spain}
\affil[3]{EHU Quantum Center, University of the Basque Country (UPV/EHU), Faculty of Science and Technology, Barrio de Sarriena s/n, Leioa, 48940, Spain}
\begin{document}
\maketitle
\begin{abstract}
Quantum key distribution (QKD) enables unconditionally secure symmetric key exchange between parties. However, terrestrial fibre-optic links face inherent distance constraints due to quantum signal degradation. Traditional solutions to overcome these limits rely on trusted relay nodes, which perform intermediate re-encryption of keys using one-time pad (OTP) encryption. This approach, however, exposes keys as plaintext at each relay, requiring significant trust and stringent security controls at every intermediate node. These “trusted” relays become a security liability if compromised.

To address this issue, we propose a zero-trust relay design that applies fully homomorphic encryption (FHE) to perform intermediate OTP re-encryption without exposing plaintext keys, effectively mitigating the risks associated with potentially compromised or malicious relay nodes. Additionally, the architecture enhances crypto-agility by incorporating external quantum random number generators, thus decoupling key generation from specific QKD hardware and reducing vulnerabilities tied to embedded key-generation modules.

The solution is designed with the existing European Telecommunication Standards Institute (ETSI) QKD standards in mind, enabling straightforward integration into current infrastructures. Its feasibility has been successfully demonstrated through a hybrid network setup combining simulated and commercially available QKD equipment. The proposed zero-trust architecture thus significantly advances the scalability and practical security of large-scale QKD networks, greatly reducing reliance on fully trusted infrastructure.
\end{abstract}
\section{Introduction}
\label{sec:introduction}
Quantum key distribution (QKD) is a cryptographic mechanism that leverages the principles of quantum physics to ensure an unconditionally secure process for key generation and exchange \cite{Shannon1949}. The key generation process is generally carried out by a quantum random number generator (QRNG) embedded within the QKD modules. This component provides perfect entropy, ensuring that the keys are completely unpredictable regardless of the computational power of the attacker. Once generated, the keys are encoded into quantum states using photon polarization. The exchange process is safeguarded against eavesdropping through the uncertainty principle and the no-cloning theorem, which make it impossible to intercept messages without irreversibly altering them.

One of the main challenges for the mass adoption of QKD is the scalability of the solution in terrestrial fibre optics networks \cite{Rajpoot2023}. Most QKD systems only guarantee a secure key exchange between point-to-point links over optical fibre. In addition, such deployments are constrained by the distance limitations of current optical technology. Because quantum channels are highly sensitive to medium imperfections, direct links beyond 300 km are difficult to achieve \cite{Louis2010}\cite{Korzh2015}\cite{Brassard2000}\cite{Rosenberg2009}.

To extend the use of the service in classical metropolitan telecommunications networks, in 2008, the secure communication based on quantum cryptography (SECOQC) network in Vienna was the first to establish a network of 85 km by interconnecting seven quantum links via optical fibre with devices of different technologies and manufacturers \cite{alleaume2007}. The SECOQC network introduced the concept of a trusted node, also known as trusted relay, as an intermediate element between two quantum links. The trusted node can receive the encrypted initial key from the previous segment, decrypt it first, encrypt it again with the quantum key of the next hop, and route it to the final recipient. This approach extends service coverage and ensures compatibility across devices from multiple vendors \cite{Louis2010}.

Various strategies for implementing this functionality are outlined in \cite{Dervisevic2024}, where the authors analyse different approaches to key management in prominent QKD networks from the past. A notable commonality across all the methods discussed is that, starting from the second hop, the original key must be reverted to plaintext before being re-encrypted with the next key in the chain. This process exposes the key to the system, making it potentially accessible to node administrators or third parties. Therefore, it becomes essential for all network participants to trust that these systems are effectively safeguarded against both internal and external threats.

Current technological solutions aimed at overcoming this limitation are either not yet commercially mature or involve specialized hardware significantly more expensive than conventional QKD protocol-based commercial equipment. Among the most promising physical-layer approaches are quantum repeaters \cite{Briegel1998}, devices designed to extend quantum communication range by preserving quantum state entanglement across multiple segments without intermediate measurements, thus maintaining end-to-end quantum security guarantees \cite{Nicolo2015}. However, quantum repeaters are still in early stages of practical development and remain far from commercially viable deployments \cite{Huttner2022}. Another noteworthy approach involves innovative protocols such as Twin-Field QKD (TF-QKD) and Measurement-Device-Independent QKD (MDI-QKD) \cite{Mohit2024}, both designed to address limitations in traditional QKD systems. TF-QKD increases effective transmission distance by exploiting single-photon interference from distant transmitters at a central node, significantly mitigating channel loss \cite{Nilesh2024}. In contrast, MDI-QKD eliminates vulnerabilities associated with detector attacks by centralizing photon measurements in an untrusted intermediate node \cite{Nicolo2015}. While these protocols represent meaningful advancements over conventional prepare-and-measure or entanglement-based schemes, their broader deployment faces challenges related to network-wide technological integration and limited commercial availability.

Until solutions like some of the previously mentioned become the state-of-the-art, the security of QKD networks will continue depending on the reliability of the intermediate elements acting as classical relays between links. At the moment, these elements represent the most vulnerable point in the security of these networks \cite{Lella2023}. Therefore, reducing trust assumptions about these network elements represents a significant step forward and one of the most relevant challenges in deploying this technology.

In this manuscript, we address the challenge of significantly reducing the trust requirements that QKD network members place on relay-node administration and security, enabling QKD networks comprised of several parties that could be owned by states or private organisations that might not necessarily comply with the same interests but share a common need to be part of a distributed secure key exchange infrastructure. It is important to distinguish between research on quantum repeaters or any other physics technological approach and classic relays used in QKD networks. This research work is focused on the latter. At the same time, the proposed methodology ensures affordable technical requirements to make the solution suitable for standard equipment. As an additional value, it also contemplates methods to leverage the confidentiality level of the process by incorporating specialised cryptographic hardware and confidential computing resources. 

The solution has been implemented with the latest European Telecommunications Standards Institute (ETSI) ISG QKD interface standards in mind, such as ETSI-014 \cite{ETSI014} for key exchange and the upcoming ETSI-020 \cite{ETSI020} for Key Management System (KMS) interoperability. The testbed comprises three simulated quantum links based on the prepare and measure schema and a physical deployment of nearly 1 km length of dark fibre between two TECNALIA facilities in Biscay, Spain. The relevance of including an actual QKD deployment with commercial equipment in the testbed is about demonstrating how this method can effectively perform with state-of-the-art devices and not only simulating their interfaces. In addition, \textbf{the methodology described in this work has been registered as patent request in the European Patent Office under the application number EP24383397.7 by Tecnalia Research \& Innovation and the University of the Basque Country.}

This paper presents the following contributions beyond the state-of-the-art:
\begin{itemize}
    \item \textbf{An enhanced confidentiality protocol among QKD network members} that ensures trust in relay node operators during all operational phases (node deployment and operation) through the use of fully homomorphic cryptography schemas.
    \item \textbf{A more resilient infrastructure against QRNG security flaws} detaching the consumer keys delivered to secure application entities (SAEs) from the QKD internal QRNGs while providing flexibility to replace the source cryptographic device.
    \item \textbf{An extra method to enhance confidentiality} by integrating specialized cryptographic hardware, while preserving core functionality within cost-effective, practical requirements
\end{itemize}

This work is structured as follows: In section \ref{sec:related-work}, we present an overview of other remarkable papers addressing the key confidentiality challenge in QKD networks using different approaches. In section \ref{sec:current-qkdn}, the current QKD network (QKDN) layered model is briefly recapped, followed by section \ref{sec:proposed-approach} with a detailed description of the method and relay architecture that we propose. Section \ref{sec:experimental-setup} is focused on depicting the experimental setup built to demonstrate the proposed method and how the approach has been designed with the current QKD industry standardisation trends in mind. Next, in section \ref{sec:results-and-discussion}, we provide some performance indicators of the solution and a theoretical explanation of the process with some thoughts about possible issues to have in mind or potential improvements. Finally, section \ref{sec:conclusions} summarizes the benefits of our method and outlines future research directions.

\section{Related Work}
\label{sec:related-work}
A key challenge for researchers seeking to expand QKD network coverage without compromising security is removing trust assumptions on relay nodes. One of the first approaches to solving this issue was proposed by Schartner et al. \cite{Schartner2009}. The authors, with access to SECOQC network in Vienna introduced in section \ref{sec:introduction}, evaluated the risk taken by all network members when using trusted nodes to relay the keys between separated quantum links. The proposed method enhances the security of trusted relay nodes in QKD networks by combining input and output keys as soon as they are established and immediately deleting the original keys. This prevents an attacker, even if they compromise the node, from accessing past or future keys. The technique reduces key storage requirements and delays the compromise of transmitted messages, allowing time for intrusion detection and response before security is fully breached.

The approach applies to two types of nodes: Quantum Access Nodes (QANs) acting as relay nodes and QANs acting as routing nodes. Relay nodes serve as classical repeaters to re-encrypt keys before forwarding them through the network, making them vulnerable if compromised. The authors secure them by ensuring that no key exists in its original form after re-encryption. Routing nodes, which handle multiple connections, require a variation of the approach to prevent keys from being reused across different paths, avoiding potential leaks if one link is compromised. Although this improves resilience, it does not fully eliminate the need for trusted nodes and relies on timely attack detection.

Other works like \cite{Lella2023} perform a very rigorous analysis of how the current QKDN architectures recommended by standardisation bodies like the International Telecommunication Union (ITU) can only improve their security by using post-quantum cryptography (PQC) algorithms for enabling a solid authentication mechanism among the members of a network. The authors propose enhancing key forwarding confidentiality in QKDNs by introducing a centralized key management agent (KMA), ensuring only the source and destination nodes must be fully trusted. A random key is generated at the source, transformed using XOR operations, and relayed through intermediate nodes without exposing the original key. This approach prevents compromised relay nodes from revealing the secret key while maintaining efficiency. However, the centralized KMA acts as a single point of failure, meaning its compromise could endanger the integrity of the entire key forwarding process. 

Additionally, the authors propose integrating PQC into QKDNs to enhance authentication and key management, which remain vulnerable to quantum attacks. They suggest using PQC-based key encapsulation mechanisms (KEMs) to generate authenticated initial keys and PQC digital signatures to establish trust between nodes, ensuring quantum-safe authentication while preserving QKD security guarantees. The hybridization of PQC and QKD technologies represents a promising research direction, as both have the potential to mitigate the quantum threat posed by algorithms such as Shor’s \cite{Shor1994} and Grover’s \cite{Grover1996}.

In \cite{Lemons2023}, Lemons et al. proposes extending QKD through proxy re-encryption to allow untrusted relay nodes to facilitate secure key forwarding. The method introduces a trusted setup phase, during which relay nodes generate proxy re-encryption transforms between each pair of connected nodes. These transforms enable the relay to convert an encrypted key from one node into an encryption under another node key, without learning the secret key or plaintext message. The system leverages learning with errors (LWE)-based cryptography and homomorphic key-switching to maintain security, ensuring that relay nodes do not store raw QKD keys after the setup phase. However, some operational phases, like the setup process, still require some trust in authorised personnel to perform the initial deployment, especially for deploying private certificates of the rest of the nodes. Plus, there are practical inconveniences to trying to keep all the certificate deployments across the network synchronized in a secure manner. 

Similarly, the authors of \cite{Geitz2023} explore how to replace the classical cryptography tied to the ETSI interfaces of QKD systems for PQC alternatives and also propose a method to mask the key in transport in a trusted node simply performing a bitwise XOR operation between the QKD key ciphered with a PQC encryptor and a QKD key in Berlin OpenQKD testbed. While the first contribution clearly improves the security of the key distribution within the trusted perimeter, the proposed masking method faces practical difficulties due to the variable and usually long length size of the produced PQC ciphertext and the typical range of QKD produced key lengths supported by the QKD modules. 

Recent works, such as \cite{Vyas2024}, propose alternative methods for preserving the confidentiality of QKD keys within a network. Specifically, the authors introduce two key management approaches. The first relies on partially trusted relays, employing multipath transmission and secret sharing to distribute key fragments across distinct paths. The final key is reconstructed at the destination, ensuring that no single relay possesses complete access to the secret key. The second approach involves a centralised KMS, wherein intermediate relays never access the original key. Instead, they perform XOR operations on locally shared QKD keys, and the central KMS aggregates these masked values to reconstruct the key at the destination.

While this approach effectively mitigates trust assumptions on relay nodes, its technical implementation may be complex and difficult to scale, particularly in networks subject to frequent structural changes. Additionally, it introduces challenges related to multi-party processes, including distributed configuration management, peer availability, and network robustness.



James et al. \cite{James2023} suggests another hybrid key forwarding mechanism in QKDNs, combining QKD and PQC to improve key confidentiality in trusted nodes. The method derives a final key from two independent sources: a QKD-generated key with Information-Theoretic Security (ITS)\cite{Shannon1949} properties and a PQC-based KEM key, ensuring end-to-end security. These two keys are combined using an XOR operation, ensuring that even if QKD or PQC is compromised, the final key remains secure. However, intermediate nodes still see the QKD key in plaintext before combination, which could be exploited if a trusted node is compromised. 

Wrapping up, a summary of the benefits and drawbacks that we identified in the discussed works are represented in Table \ref{tab:works_comparison}.

\begin{table}[!h]
    \centering
    \begin{tabular}{@{}llp{3cm}p{6cm}l@{}}\toprule
        \textbf{Authors} & \textbf{Architecture} & \textbf{Complexity} & \textbf{Observations} \\
        \midrule
        Schartner et al. \cite{Schartner2009} & Intermediate nodes & High (requires different key management roles and active monitoring)& Does not completely eliminate the need for trusted nodes, security relies on early intrusion detection, and there is a risk of attacks based on collecting newly generated keys.\\
        Lella et al. \cite{Lella2023} & Centralised KMA & Medium (requires complex network topology) & Depends on a highly available and secure centralized KMA. May add complexity to routing and network management.\\
        Lemons et al. \cite{Lemons2023} & Intermediate nodes & High (requires manual certificate management) & The trusted setup phase is a single point of failure, if a relay is compromised before key deletion, security is lost. Additionally the private key synchronization among network members can be challenging.\\
        Geitz et al. \cite{Geitz2023} & Intermediate nodes & Medium (requires PQC integration in standard interfaces) & Assume that PQC can be integrated in all components of the network. Therefore, the variable length of PQC ciphertexts can difficult OTP operations.\\
        Vyas et al. \cite{Vyas2024} & Hybrid & High (requires complex network administration) & The centralised KMS is a single point of failure, its compromise results in total security loss. In the partially trusted approach demands precise key synchronization across multiple paths, adding complexity to the network maintenance.\\
        James et al. \cite{James2023} & Hybrid & Medium (requires extending standard application interfaces) & The proposed methodology requires to extend current QKD interface standards and requires PQC integration in many components of a network. \\
        \bottomrule
    \end{tabular}
    \caption{Summary of different approaches for reducing trust on QKD relay nodes.}
    \label{tab:works_comparison}
\end{table}

This manuscript focuses on significantly reducing trust assumptions in QKD relays while transmitting keys over classical channels. The proposed approach preserves the state-of-the-art one-time pad (OTP) method, ensuring that the ITS nature of key forwarding remains intact while adding an extra confidentiality layer to local key processing within each node. This is achieved without introducing excessive complexity or requiring a trusted setup phase. The method is compatible with any network routing strategy and remains independent of the underlying QKD protocols or device models, allowing for potential enhancements in local node confidentiality when specific hardware is available.

\section{Current QKDN architecture}
\label{sec:current-qkdn}
Before describing the confidentiality-increased architecture that we propose, it is important to briefly recap how the current state-of-the-art QKD networks are usually organised. This will help the reader understand the concerns about trust and how this work tries to solve this challenge. For clarity, all graphical examples in this work assume a simple network with a limited number of quantum links. The routing methods within the network will also be out of the scope. The solution proposed in this manuscript is fully compatible with any routing strategy. In fact, routing is a very mature research field full of high-quality solutions already used in production networks \cite{Martin2024}.

\begin{figure}[ht]
    \centering
    \includegraphics[width=1\linewidth]{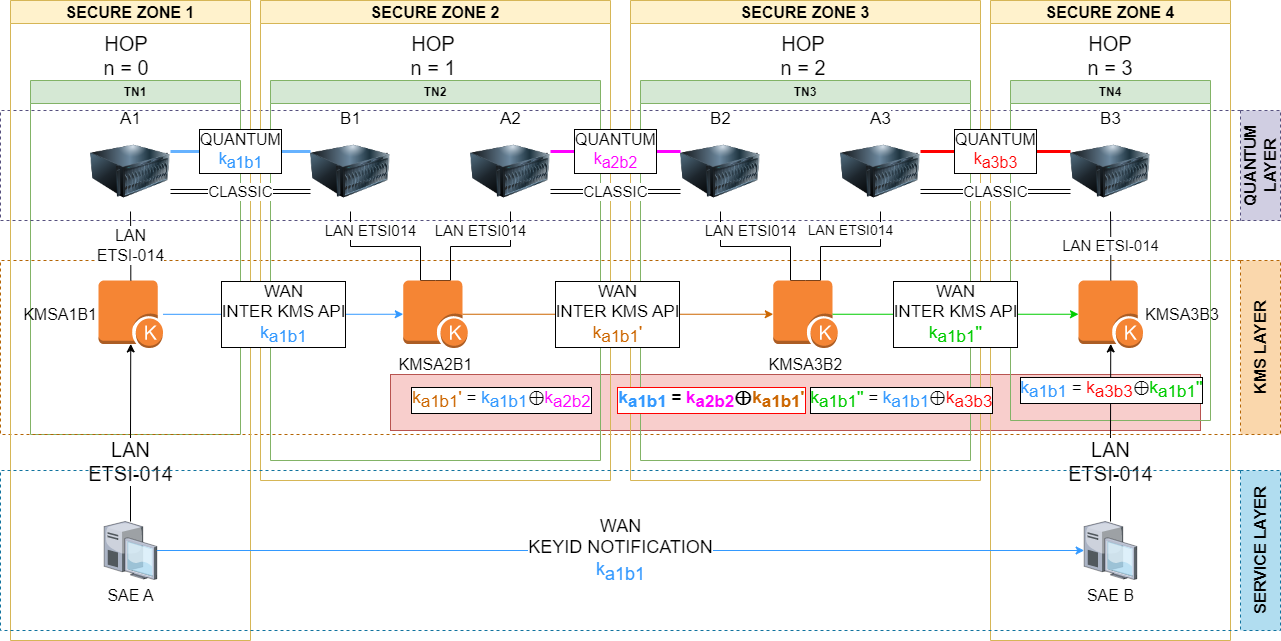}
    \caption{Current QKD networks architecture based on intermediate nodes}
    \label{fig:qkdn-vanilla}
\end{figure}

The current model for exchanging keys in a QKD network is depicted in Fig. \ref{fig:qkdn-vanilla}. There are other topologies based on a centralised KMS common to all nodes, but the core operation to forward the key among trusted nodes falls under the same principle. The network is vertically organised as follows, from top to bottom:
\begin{enumerate}
    \item \textbf{Quantum layer:} This is the plane where the QKD appliances operate. Each pair of prepare-and-measure emitter and receiver is connected by its own optics fibre and the additional classic wiring it needs to run based on the manufacturer's specifications. As of the time of writing, the interoperability standard ETSI-020 \cite{ETSI020} between KMS has not yet been published and remains in draft form. Each QKD device, therefore, has its own KMS, which we will refer to as Key Management Entity (KME), and only provides a downstream interface to deliver the keys service based on ETSI-014 \cite{ETSI014}.
    \item \textbf{Key management layer:} This is the layer where the local KMS of each trusted node handles key forwarding and provides service to end users. The KMS is a specialized software running on a conventional computer that acts as a key consumer from the KMEs of the QKD vendor. This occurs whenever a service layer end-user requests to share a key with another user.
    In particular, if the KMS is not situated at the edge of the network, it connects to at least two different quantum Links. It securely houses one QKD appliance for each link it is connected to. Once the vendor KMEs implement the KMS interoperability application programming interface (API), the external KMS will no longer be necessary and the QKD-integrated KMS will then be able to perform the same operations without relying on this external entity for key consumption.
    \item \textbf{Service layer:} in this layer, the final key consumers, also known as SAE, perform cryptographic operations with the keys provided by the QKD network and notify them of the key IDs used to encrypt their information. The internal network topology of previous layers is completely transparent, and they can be connected by classical internet protocol (IP) networks, typically on a wide area network (WAN).
\end{enumerate}

The previously depicted schema, the security of keys in transit between nodes is fully ensured by the OTP encryption based on XOR operations with QKD keys of the same length. However, concerns arise regarding the confidentiality of keys within each relay, particularly if a trusted node is compromised, allowing a malicious actor to expose all keys routed through it. Additional challenges in this model include the lack of crypto-agility. This means that being tied to the embedded QRNG components of the QKD modules makes it hard to replace the random source device in case of detecting any security flaw by one vendor. Furthermore, using the same QRNG-generated QKD material as the final SAE keys pushes the first OTP encryption operation up to the second hop, revealing the initial key to the first trusted node without reversing any XOR operation. Lastly, the high trustworthiness levels assumed by all network members across all operational phases on node administration and safeguarding operations from setup to maintenance activities.

\section{Proposed Approach: Zero-Trust QKD Relay}
\label{sec:proposed-approach}
During the last decade, organisations have moved towards zero-trust architectures (ZTA) in their internal information processes to avoid assuming trust at any point in the system. As defined by the National Institute of Standards and Technology (NIST) \cite{NIST2020}, zero trust architecture (ZTA) shifts security enforcement from a perimeter-based model towards validating individual transactions independently. Among its key design principles is the minimization of trust boundaries within the system, enforcing security policies as closely as possible to the actual resources being protected, rather than relying on broad trust zones. This approach not only provides better security, moving policy enforcement closer to resources but also provides more flexibility for all members of an organisation. When there is no perimeter anymore, the involved agents enjoy more freedom to perform their roles with a variety of hardware and location combinations if they comply with the baseline policy set. Although ZTA covers a broader set of considerations in information systems security, some concepts, like reducing the trust surface on all actors involved in a process, can be applied to improve the current state-of-the-art of QKD networks. 

Fully homomorphic encryption (FHE) schemes and specialized cryptographic hardware for confidential computing are key technologies designed to address the trust requirements in relay QKD nodes. These technologies provide a practical solution for ensuring confidentiality in existing networks without requiring drastic architectural changes. FHE cryptographically enables bit-by-bit OTP operations without exposing the original keys in plaintext to the processing node, thereby eliminating trust assumptions about node administrators. Performing the same OTP operation with a QKD transmitted private key, this method adds an additional confidentiality layer to the operation in the local key processing without interfering in the secure key forwarding ITS nature.

As LWE is currently considered part of the quantum-resistant family of cryptographic algorithms, it provides a strong confidentiality layer to the existing OTP cipher used between quantum-generated keys. Additionally, the approach presented in this paper further enhances security by leveraging specialized cryptographic computing resources to conceal homomorphic operations, even from the hardware executing the process.

In this approach, no one knows the quantum keys except the SAEs involved in the exchange process between themselves and the local relays providing the service in their secure zone. 

\subsection{Detaching QKD keys from the initial secret}
In the traditional approach, the same keys generated by the QKD modules play two different roles: as the secrets provided to the end users of the service and the material to perform the OTP encryption between nodes. In the proposed zero-trust architecture, to keep the consumer keys visibility restricted to the secure zone perimeter where SAE-A initiates the process, each relay-KMS will be equipped with an external QRNG \cite{Saini2022} to generate the secrets delivered to every SAE request in its zone. 

The addition detaches the QKD internal QRNG module keys from the keys consumed by end users of the QKDN service. Furthermore, this modification enhances the crypto agility of the system in cases where a QRNG used by a relay-KMS is found to be vulnerable. The modular design facilitates replacing a compromised QRNG with another unaffected one without significantly impacting the architecture. In contrast, if the embedded QRNG module of a commercial equipment is found vulnerable, the whole QKD device must be replaced, since vendors usually do not allow customers to manipulate internal parts. 

However, including an additional QRNG in the relay-KMS introduces certain implications related to managing the lifecycle of the associated key. Since the key is generated externally to QKD KMS modules, these modules remain unaware of its existence. Consequently, every QRNG key created by the relay-KMS to meet a SAE demand must be securely buffered in a pool of activated keys until the destination SAE requests it. To achieve this, the relay-KMS generates a Key ID for each key and delivers it to the SAEs in accordance with the ETSI-014 API reference. Another notable advantage is that the key obtained by SAE-A does not originate from a QKD module, enabling the first hop in the path to perform an XOR operation with the QKD key generated in the initial segment.



At this point, we have a scenario in which the keys used by SAEs are different from the QKD segments. QKD keys are only used to perform OTP encryption between hops. However, each relay still unveils the original key when undoing the XOR with the previous segment. To avoid this from happening, all relay members of the network operate the OTP encryption under a FHE schema based on Brakerski/Fan-Vercauteren scheme (BFV) \cite{Brakerski2012} \cite{Fan2012}.

\subsection{Confidential key processing with FHE}
\label{confidential}
Homomorphic encryption (HE) is a set of techniques that allows certain arithmetical operations to be performed directly over encrypted data. This enables a group of untrusted members to operate over a piece of information encrypted with a public key while keeping the ciphertext exclusively available for decryption to the owner of the related private key. Among the existing cryptographic techniques within this family, the election criteria must meet the highest protection level with the quantum threat in mind.

In FHE, there are different cryptosystems to choose from based on the kind of mathematical problem on which its security is based or the required operations to perform with ciphertexts. The choice of the BFV scheme for this architecture is primarily motivated by the robustness of its underlying algorithm, which is based on the LWE problem. This family of cryptosystems is currently considered \textbf{quantum resistant} \cite{Regev2009} since there is no quantum algorithm today that can weaken its properties. Other alternatives, such as the Brakerski-Gentry-Vaikuntanathan (BGV) scheme, also possess similar properties and could have been suitable as well. Moreover, since all FHE schemes rely on hard lattice-based problems, their security levels are comparable, meaning that if a vulnerability were found in one of them, migrating to another unaffected scheme would be necessary. However, in such a case, the key forwarding process would not compromise security, as the key in transit between QKMS remains encrypted with OTP. A relevant example is ML-KEM Kyber\cite{Kyber}, recently standardized as Federal Information Processing Standard (FIPS) 203 by NIST\cite{NIST2024}, which is based on a variant of LWE. The fact that it has undergone extensive scrutiny by the cryptographic community further reinforces the confidence in the hardness of LWE-based schemes. In addition, being part of the group of FHE algorithms provides enough versatility to change the operation performed in the OTP process in the future. Another positive point is its good performance in some operations with integers and the noise control in the process, thanks to its levelled-based public key ciphering.

The working principle closely mirrors classical asymmetric cryptography. Each participant possesses a public-private key pair, and the sender encrypts data using the public key of the recipient. However, while the ciphertext is encrypted for the destination node with its public key, its contents consist of the initial key XORed with the next QKD key link in the path. This approach does not replace the OTP ITS mechanism for key encryption but instead preserves the OTP process, ensuring that the key remains encrypted at the node responsible for re-encrypting it with the key of the next QKD link.

Despite these similarities, the scheme differs in two fundamental aspects. First, it enables not only encryption and decryption but also operations directly on encrypted content. Second, any entity that intends to operate on the ciphertexts must configure the same cryptographic context parameters as those defined by the owner of the private key. Considering this, the next challenge to address is the XOR operation required by the relay-KMS during the key-forwarding step, as illustrated in Table \ref{tab:xor_gate}.

\begin{table}[!h]
    \centering
    \begin{tabular}{@{}lcccl@{}}\toprule
        \textbf{A} & \textbf{B} & \textbf{$A \oplus B$} \\\midrule
        0 & 0 & 0 \\
        0 & 1 & 1 \\
        1 & 0 & 1 \\
        1 & 1 & 0 \\\bottomrule
    \end{tabular}
    \caption{XOR gate}
    \label{tab:xor_gate}
\end{table}

To reproduce the same behaviour operating in the homomorphic plane of BFV, the operation can be represented as an addition operation in the Galois Field (GF) of order 2 ($\texttt{GF}(2)$). This operation is arithmetically equivalent to the XOR gate. Recall that, since $\texttt{GF}(2)=\{0, 1\}$, the arithmetical addition operation is always reduced modulo 2, as shown in Table \ref{tab:xor_equivalent_homomorphic}.

\begin{table}[!h]
    \centering
    \begin{tabular}{@{}lccl@{}}\toprule
        \textbf{A} & \textbf{B} & \textbf{A + B}\\\midrule
        0 & 0 & \(0 \bmod{2}=0\)\\
        0 & 1 & \(1 \bmod{2}=1\)\\
        1 & 0 & \(1 \bmod{2}=1\)\\
        1 & 1 & \(2 \bmod{2}=0\) \\\bottomrule
    \end{tabular}
    \caption{XOR equivalent operation in addition operation under GF2}
    \label{tab:xor_equivalent_homomorphic}
\end{table}

Using only addition offers a significant advantage. Compared to multiplication, the noise generated during this process is much lower. This allows the system to perform homomorphic XOR operations on encrypted keys numerous times within large networks without increasing the noise to unacceptable levels. More details on how this parameter is affected on each subsequent operation will be discussed in Section \ref{sec:results-and-discussion}.

Implementing this process to bit strings of length 256 brings another challenge. Adding two keys of this size in GF2 provokes overflows or unintended carries when performing arithmetic operations. A carry is generated in binary addition when the sum of two bits exceeds the maximum value that can be represented in a single bit. In the simplest case, adding (\(0 + 0)\) results in 0, and 0 + 1 or 1 + 0 results in 1, which fits within a single bit. However, when adding 1 + 1, the result is 10 in binary (which is equivalent to 2 in decimal). Since a single bit can only store 0 or 1, the 0 remains in the current bit position, while the extra 1 (the carry) is propagated to the next higher bit.

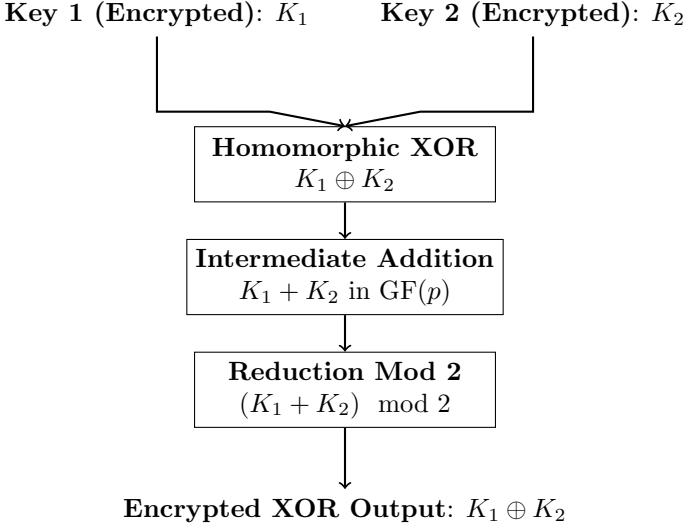
\begin{figure}[h]
    \centering
    \adjustbox{max width=\textwidth}{ 
    \begin{tikzpicture}[node distance=2cm, scale=1, every node/.style={scale=1}]
        \node (k1) at (-1.5,5.8) {\textbf{Key 1 (Encrypted)}: $K_1$};
        \node (k2) at (3.5,5.8) {\textbf{Key 2 (Encrypted)}: $K_2$};

        \node[draw, rectangle, minimum width=4cm, minimum height=1cm, align=center] (xor1) at (1,3.8) {\textbf{Homomorphic XOR}\\$K_1 \oplus K_2$};

        \node[draw, rectangle, minimum width=4cm, minimum height=1cm, align=center] (add) at (1,2.3) {\textbf{Intermediate Addition}\\$K_1 + K_2$ in GF($p$)};

        \node[draw, rectangle, minimum width=4cm, minimum height=1cm, align=center] (mod2) at (1,0.8) {\textbf{Reduction Mod 2}\\$(K_1 + K_2) \mod 2$};

        \node (output) at (1,-0.8) {\textbf{Encrypted XOR Output}: $K_1 \oplus K_2$};

        \draw[->, thick] (k1.south) -- ++(0,-1) -- (0,4.5) -- (xor1.north);
        \draw[->, thick] (k2.south) -- ++(0,-1) -- (2,4.5) -- (xor1.north);
        \draw[->, thick] (xor1.south) -- (add.north);
        \draw[->, thick] (add.south) -- (mod2.north);
        \draw[->, thick] (mod2.south) -- (output.north);
    \end{tikzpicture}
    } 
    \caption{Homomorphic XOR operation in the BFV scheme using addition in GF(2) and final modulo 2 reduction.}
    \label{fig:homomorphic_xor}
\end{figure}

In the context of homomorphic encryption, unintended carries can alter the expected binary structure when working in a field larger than GF(2). Since XOR in GF(2) is mathematically equivalent to addition modulo 2, any carry propagation that occurs due to operating in a larger modulus must be carefully managed. By performing operations in an extended modulus and applying modular reduction at the final step, we ensure that the result retains its correct binary form without unintended overflows. Fig. \ref{fig:homomorphic_xor} represents graphically the steps mentioned. In the figure, \(p\) in \(GF(p)\) represents the larger prime field where intermediate homomorphic additions are performed before applying the final modulo 2 reduction to ensure binary consistency. 

Finally, when operating in a modulus larger than 2, intermediate values can exceed the binary range {0,1}, leading to an expanded representation that requires additional bits. Even after applying a final modulus 2 reduction to restore binary values, the computation may have temporarily used a larger space, resulting in a longer output than the original input. This happens because, in a higher modulus (e.g., 16), sum operations can produce values requiring multiple bits (e.g., up to 4 bits for values between 0 and 15), which are retained until the reduction step. Once reduced modulo 2, the output is binary again, but some extra bits may remain from intermediate calculations. To ensure the final result maintains the original key length, these excess bits must be truncated to 0, preventing unintended data artifacts and preserving the structure necessary for correct decryption.

To better understand the above explanation, there is a simple example represented in Fig.  \ref{fig:overflow-example} where two-bit strings with binary values are operated in a bigger space, obtaining values that exceed the two possible binary values as an intermediate result before reducing the outcome in a smaller space.

\begin{figure}[!ht]
\begin{verbatim}
String 1:  1 0 1 1
String 2:  1 1 0 1
---------------------
Int. result (mod 16):  2 1 1 2
Int. result (binary): 0010 0001 0001 0010
---------------------
Apply mod 2: 0000 0001 0001 0000
Result truncating length to GF(2):  0 1 1 0
\end{verbatim}
\caption{Intermediate operations in larger spaces}
\label{fig:overflow-example}
\end{figure}

Performing intermediate operations in larger spaces and applying the reduction only to the final result helps avoid losing information by prematurely reducing excessive values. This approach preserves precision in the results, especially when chaining multiple operations. Additionally, it minimizes the noise introduced by these operations by avoiding intermediate reduction steps \cite{Brakerski2012}.

In the tests implemented during the investigation, this method produced a solid XOR operation in a homomorphic plane while keeping the ciphertext to a controlled size and being consistently precise, enabling the recovery of the initial key value with no errors.

\subsection{Optional cryptographic hardware}
Externally generated QRNG keys and OTP encryption with QKD keys under FHE already guarantee not only the same ITS protection to the key in transit typical from QKD. It also offers an unprecedented level of confidentiality to SAEs in QKD networks, concealing the keys exchanged across the network from the intermediate relay and ensuring that they do not have access to the keys they forward. However, the current state-of-the-art cryptographic hardware provides extra tools to increase the isolation of the homomorphic key operations.

A trusted execution environment (TEE) is a secure area within a device processor that prevents the data and processing performed within it from being accessible to the rest of the system \cite{Sabt2015}. In other words, it consists of an area of the hardware isolated from the rest of the components that allow the processing of information only by authorized code, guaranteeing confidentiality with respect to the system on which it is running.

Different types of hardware allow execution in a TEE environment, depending on the manufacturer and the architecture. Most are implemented as a particular set of instructions to handle specific operations in a reserved area of memory completely isolated from the rest of the system. This solution can be run on its special hardware or as a confidential computing service that some cloud providers offer \cite{AzureConfidential}. This extra layer adds another level of confidentiality to the key in transport from potential eavesdroppers with system-level access.

Also, mention the possibility of storing the relay homomorphic private key in a trusted platform module (TPM) \cite{Tomlinson2017}. This component, already present in many commercial hardware, can help to confidently store the private key securely, preventing a situation where a third party would be able to break into the system and steal the keys from some members of the network. 

\subsection{All pieces together}
With this method, the keys used by the end users of the network keep their QKD ITS condition while, at the same time, none of the intermediate QKD relays participating in its transport can see them. To put all the pieces together, a QKD network comprised of the presented model of zero-trust QKD relays (ZTQR) as represented in Fig.  \ref{fig:qkdn-ztqr} will streamline its process in the following way:

\begin{figure}[!ht]
    \centering
    \includegraphics[width=1\linewidth]{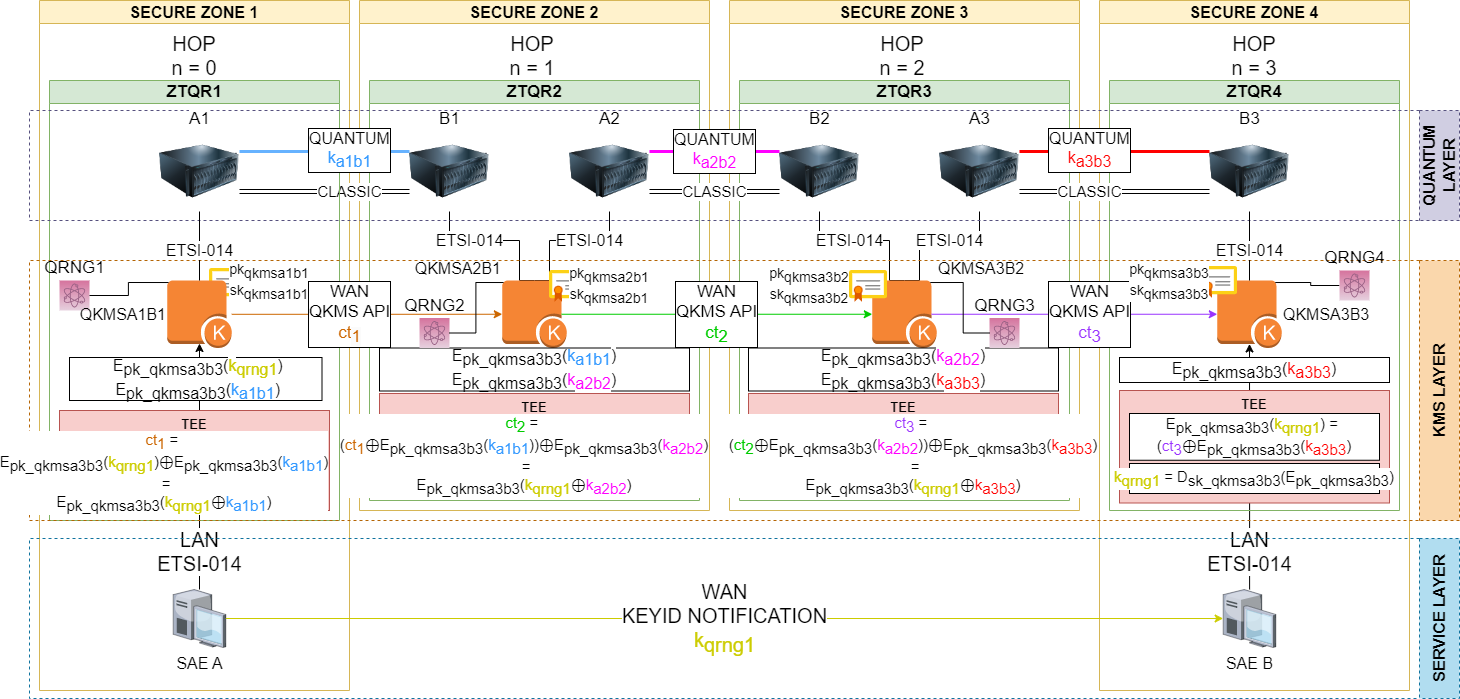}
    \caption{Zero-trust QKD relay based QKDN}
    \label{fig:qkdn-ztqr}
\end{figure}

\begin{enumerate}
    \item Each of the ZTQR generates, if there is not already, a pair of private and public homomorphic keys and a cryptocontext on their startup process. The public key \(qkms_{pk}\) and the cryptocontext are serialized to a file and placed in a network-wide accessible repository, ready to be used by other network members (the implementation of this sharing space is out of the scope of the presented method). The private counterpart \(sk_{qkms}\) is generated and stored privately in its filesystem or a specific cryptographic chip.
    \item Whenever a SAE requests to share a key with another located in another secure zone, it requests for a key to exchange using the ETSI-014 API to its local ZTQR. In this case, SAE-A wants to share a key with SAE-B.
    \item ZTQR1, which received the request of SAE-A generates a new quantum key with its own QRNG in response to SAE-A demand (\(k_{qrng1}\)). At the same time, it checks the destination ZTQR (in this example, ZTQR4) and requests another key from the first QKD link in the chain to reach this destination (\(k_{a1b1}\)).
    \item ZTQR1 encrypts \(k_{qrng1}\) and \(k_{a1b1}\) with the public key of the destination ZTQR4 and performs an XOR in the homomorphic plane generating a ciphertext (Eq. \ref{eq:encrypt-pk}). 
    \begin{equation}
        ct_{1} = E_{qkmsa3b3_{pk}}(K_{qrng1}) \oplus E_{qkmsa3b3_{pk}}(K_{a1b1}) = E_{qkmsa3b3_{pk}}(K_{qrng1} \oplus K_{a1b1})
        \label{eq:encrypt-pk}
    \end{equation}
    \item ZTQR1 sends (\(ct_{1}\)) over a classic channel in a "\/ext\_keys" ETSI-020 compliant payload to ZTQR2, which is the next in the path to reach ZTQR4
    \item ZTQR2 receives the "\/ext\_keys" message from ZTQR1. Among the data on the payload, there is the destination SAE to deliver the key. If it is not within its perimeter, the next hop will be calculated based on the network's routing algorithm. But first, it has to undo the XOR on the ciphertext with \(k_{a1b1}\), retrieving the key from the ETSI-014 interface of the QKD belonging to the quantum Link shared with ZTQR1. As the destination SAE to reach with that key is not present, it has to perform a new XOR with a key from the next quantum Link in the chain, named \(k_{a2b2}\) (Eq. \ref{eq:reencrypt}).
    \begin{equation}
        ct_{2} = (ct_{1} \oplus E_{qkmsa3b3_{pk}}(K_{a1b1})) \oplus E_{qkmsa3b3_{pk}}(K_{a2b2}) = E_{qkmsa3b3_{pk}}(K_{qrng1} \oplus K_{a2b2})
        \label{eq:reencrypt}
    \end{equation}
    \item The process in ZTQR3 is the same as in ZTQR2 but for the last hop (Eq. \ref{eq:reencrypt-again}).
    \begin{equation}
        ct_{3} = (ct_{2} \oplus E_{qkmsa3b3_{pk}}(K_{a2b2})) \oplus E_{qkmsa3b3_{pk}}(K_{a3b3}) = E_{qkmsa3b3_{pk}}(K_{qrng1} \oplus K_{a3b3})
        \label{eq:reencrypt-again}
    \end{equation}
    \item When the key container arrives at ZTQR4, as SAE-B is within its perimeter and the ciphertext is built using its own Public Key, the only operation that has to be performed is undoing the XOR (Eq. \ref{eq:last-undo}) for the last time before decryption with the key (Eq. \ref{eq:decrypt}).
    \begin{equation}
        E_{qkmsa3b3_{pk}}(K_{qrng1}) = ct_{3} \oplus E_{qkmsa3b3_{pk}}(K_{a3b3})
        \label{eq:last-undo}
    \end{equation}
    \begin{equation}
        K_{qrng1} = D_{qkmsa3b3_{sk}}(E_{qkmsa3b3_{pk}}(K_{qrng1}))
        \label{eq:decrypt}
    \end{equation}
\end{enumerate}

In this example, both SAEs are located on the edges of the represented plain network to facilitate the process explanation. However, as every ZTQR has its own QRNG, any of them cannot only forward but also serve as a QKDN access point to exchange keys with any other node without depending on the network topology or where the SAEs involved in the key exchange are present.

\section{Experimental Setup}
\label{sec:experimental-setup}
In order to demonstrate the feasibility of the presented architecture, all ZTQR functionality and QKD KME simulations were developed using Python. The FHE functionality has been integrated with the official wrapper in Python of OpenFHE library \cite{PythonOFHE}.

OpenFHE \cite{OpenFHE} is an open-source C++ library that implements state-of-the-art quantum-resistant FHE schemes. It is well-documented, cross-platform, and backed by an active community, making it a reliable choice for cryptographic applications. Our decision to use OpenFHE is based on prior experience and its strong community support. The library also provides seamless Python integration through its official wrapper, enhancing accessibility and usability. OpenFHE supports a broad range of FHE cryptosystems, including BFV, which enables arithmetic operations on encrypted integers without decryption, ensuring data confidentiality. These capabilities make OpenFHE a powerful and versatile tool for developing advanced homomorphic encryption solutions.

\begin{figure}[!ht]
    \centering
    \includegraphics[width=0.5\linewidth]{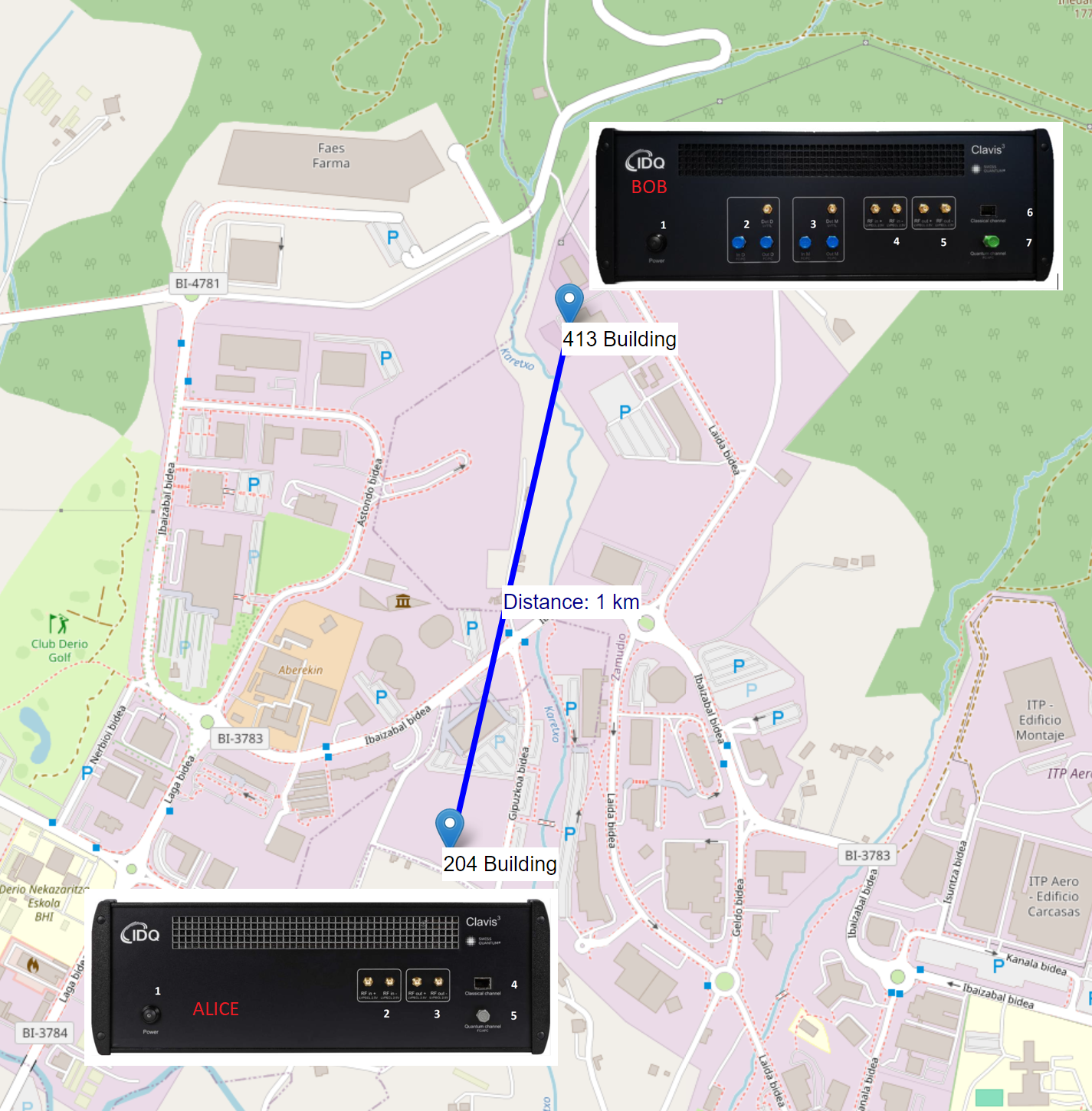}
    \caption{QKD testbed deployment in TECNALIA premises}
    \label{fig:qkd-map}
\end{figure}

The testbed is built on a hybrid QKD network using a pair of prepare-and-measure QKD appliances based on the Clavis3 platform from IDQuantique, operating under the COW3 protocol  at 1550nm \cite{Stucki2007}. The setup includes a dark fibre link of approximately one kilometer between two TECNALIA facilities in an industrial area, as shown in Fig. \ref{fig:qkd-map}. These fibres were not deployed specifically for this demonstrator but are part of the municipality existing telecommunications infrastructure.

In terms of link performance, the average QBER is around 0.03, and the key rate varies between 1900 and 2500 bits per second. While these metrics provide insight into performance, the essential requirement is that the QKD link maintains a minimum quality level that ensures proper key provisioning in the KMS buffers.

This work does not focus on the physical QKD protocol but rather on a network-level approach applicable to all QKD relay nodes that perform OTP encryption between key pairs. However, real QKD devices have been integrated to validate compatibility with the existing interface standards.

Additionally, a computer hosts Docker containers running the remaining network elements, which include five zero-trust QKD relays and six simulated QKD KMEs configured to form three QKD links. 

\begin{figure}[!ht]
    \centering
    \includegraphics[width=0.8\linewidth]{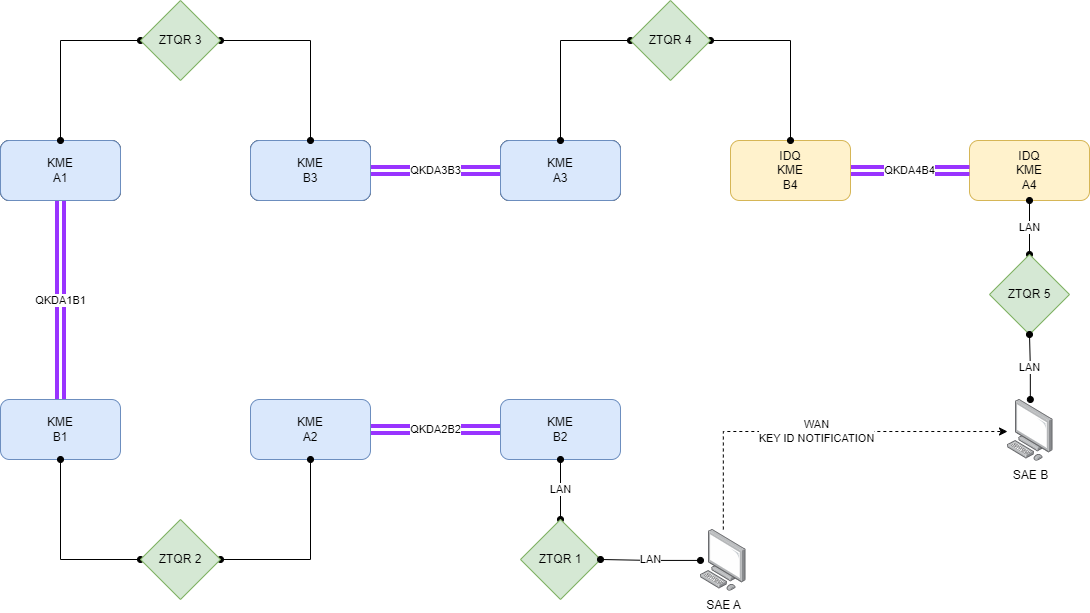}
    \caption{Zero-trust QKD network testbed}
    \label{fig:zt-qkdn-map}
\end{figure}

The detailed components shape a network topology represented in Fig.  \ref{fig:zt-qkdn-map}.

To emulate a real deployment scenario where QKDs are covering long distances linking independent LANs, a network segmentation between elements has been performed and is represented in Table \ref{tab:qkdn-network-segments}. Each secure zone has a network segment where the following elements are located: a local ZTQR, one or more QKD KME and SAEs. Each pair of QKD KMEs share a network segment as the classical channel between KMS. In addition, all ZTQR containers share an extra network adapter with another network segment, which aims to be the WAN network where the inter-KMS messages are exchanged. 

\begin{table}[!h]
    \centering
    \begin{tabular}{@{}lccl@{}}\toprule
        \textbf{Name} & \textbf{Classless Inter-Domain Routing (CIDR)}\\\midrule
        kms-link-a1-b1 & 10.1.1.0/27\\
        kms-link-a2-b2 & 10.1.2.0/27\\
        kms-link-a3-b3 & 10.1.3.0/27\\
        secure-zone-1 & 10.1.10.0/27\\
        secure-zone-2 & 10.1.20.0/27\\
        secure-zone-2 & 10.1.30.0/27\\
        secure-zone-2 & 10.1.40.0/27\\
        secure-zone-2 & 10.1.50.0/27\\
        inter-qkms-wan & 10.1.100.0/27\\\bottomrule
    \end{tabular}
    \caption{List of network segments defined in testbed}
    \label{tab:qkdn-network-segments}
\end{table}

The software process involved in the key forwarding operations starts with the initial homomorphic XOR between the first QKD and the local QRNG keys and packing it into an ETSI-020 "\/ext\_keys" message. To include the homomorphically encrypted XOR in the body of the message, the following steps are performed:

\begin{enumerate}
    \item Serialize the ciphertext to binary
    \item Compressing the binary with GNU ZIP (GZIP)
    \item Encoding the resulting GZIP in a Base64
\end{enumerate}

This process guarantees the ciphertext integrity, avoids text encoding errors in HTTP message exchanges and reduces the resulting payload size to make it compatible with most Web clients and servers.

Whenever a ZTQR receives an "ext\_keys" container within an ETSI-020 message from a previous segment, it must first unpack the encoded key using the reverse of the previously described process. The same procedure is repeated after processing the ciphertext before sending it to the next relay in the path.

\begin{figure}[!ht]
    \centering
    \includegraphics[width=1\linewidth]{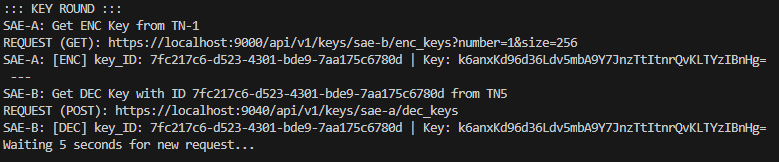}
    \caption{Test script output}
    \label{fig:test-script}
\end{figure}

To test the solution, a script was developed to get encryption keys from ZTQR1 ETSI-014 interface and recover them from ZTQR5 after all propagation events have occurred. The output of the script is shown in Fig.  \ref{fig:test-script}.

For each key round, when the destination SAE request the key, a "/void" message is propagated backwards until reaching the first ZTQR to request for the key to be deactivated in line with the current draft of ETSI-020 specification.

\section{Results and Discussion}
\label{sec:results-and-discussion}
The success of the experimentation activities carried out on the testbed demonstrated the feasibility of the architecture proposed. The solution enabled a secure and confidential method to exchange a shared secret between two parties across four quantum links with intermediate relays without comprising the initial secret to any third party involved in the process. Each node operated successfully the OTP decryption and encryption operations over homomorphic ciphertexts without any notable degradation in the service. Furthermore, the platform comply with the latest standards in QKD interfaces such as ETSI-014 and is compatible with the upcoming ETSI-020 demonstrating how the proposed solution is ready to be integrated in existing QKD networks.

In order to study how the solution scales in wider QKD networks, a crucial factor to keep under control, particularly in the context of homomorphic encryption, is the final size of the encoded ciphertext within the messages exchanged between nodes in the network. This size must remain within the limits supported by the most common HTTP servers to prevent transmission issues that could undermine the applicability of the solution in real-world environments. To mention examples of this default parameter values in some of the most popular HTTP server software:

\begin{itemize}
    \item Nginx defines this value in "ngx\_http\_code\_module" with a default value of 1MB \cite{NGXLimits}
    \item Microsoft IIS defines a default value in the "maxAllowedContentLength" attribute of 30000000 bytes (nearly 28.6 MB) \cite{IISLimits}
    \item Apache Tomcat defines a default value of "maxPostSize" of 2 MB \cite{TomcatLimits}
\end{itemize}

To obtain this value, the script, which shows its output in Fig. \ref{fig:test-script}, was executed in a one-hour loop, requesting key recovery for ZTQR-1 to ZTQR-5. The process was repeated for 128 and 256-bit keys. For each relay, time and CPU usage were measured during key processing operations (initial XOR, undo XOR, redo XOR, and final undo XOR). The size of the resulting Base64 string (after serialization and compression), processing time, and average CPU usage were recorded before being included in the POST request. It is important to note that this comparison does not evaluate two solutions providing the same level of confidentiality. Instead, it contrasts the current method, which guarantees the security of the ITS key during transport while exposing it to each intermediate node, with a proposed solution that not only maintains this guarantee but also adds an extra layer of protection to prevent intermediate nodes from accessing the key. Consequently, an overhead is expected, as the new method enhances confidentiality beyond what the baseline approach offers.

The hardware specifications of the computer used to run the tests are the following:

\begin{itemize}
    \item \textbf{OS:} Ubuntu 22.04.1 (LTS)
    \item \textbf{CPU Model:} Intel(R) Core(TM) i5-8250U CPU @ 1.60GHz
    \item \textbf{Physical Cores:} 4
    \item \textbf{Logical Cores:} 8
    \item \textbf{Total Memory:} 8 GB
\end{itemize}

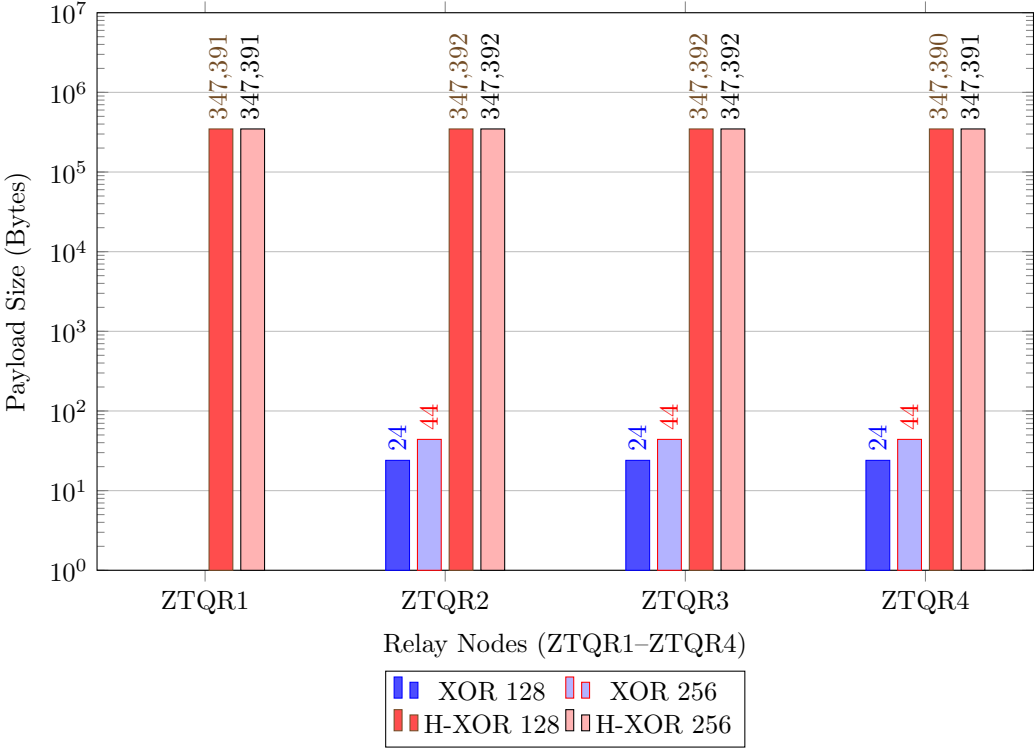
\begin{figure}[ht]
    \centering
    \pgfplotstableread[col sep=comma]{table_data_sizes.csv}\finaltable
    
    \begin{tikzpicture}
      \begin{axis}[
        ybar,
        bar width=9pt,
        width=0.9\linewidth,
        height=9cm,
        symbolic x coords={ZTQR1,ZTQR2,ZTQR3,ZTQR4},
        xtick={ZTQR1,ZTQR2,ZTQR3,ZTQR4},
        ylabel={Payload Size (Bytes)},
        xlabel={Relay Nodes (ZTQR1--ZTQR4)},
        legend style={at={(0.5,-0.18)},anchor=north,legend columns=2},
        nodes near coords,
        point meta=rawy,
        every node near coord/.append style={
            rotate=90, anchor=west,
            /pgf/number format/fixed,
            /pgf/number format/precision=0
        },
        ymajorgrids=true,
        enlarge x limits=0.15,
        ymode=log,
        log basis y=10,
        ymin=1,
        ymax=10000000,           
      ]
    
        \addplot+[fill=blue!70!white, bar shift=-18pt]
          table[x=Node,y=XorOneHundredTwentyEight]{\finaltable};
    
        \addplot+[fill=blue!30!white, bar shift=-6pt]
          table[x=Node,y=XorTwoHundredFiftySix]{\finaltable};
    
        \addplot+[fill=red!70!white, bar shift=6pt]
          table[x=Node,y=HXorOneHundredTwentyEight]{\finaltable};
    
        \addplot+[fill=red!30!white, bar shift=18pt]
          table[x=Node,y=HXorTwoHundredFiftySix]{\finaltable};
    
        \legend{XOR 128, XOR 256, H-XOR 128, H-XOR 256}
    
      \end{axis}
    \end{tikzpicture}
    \caption{Payload Size Comparison: XOR vs. H-XOR (128 \& 256 bits)}
    \label{fig:comparison_sizes}
\end{figure}

The experimental results showed no significant differences in payload sizes, as seen in Fig. \ref{fig:comparison_sizes}, regardless of key length or the number of hops. The average payload size remained stable at approximately \textbf{347 kilobytes}, demonstrating the feasibility of transmitting this amount of data within an HTTP request without issues. This is primarily because the exchanged QKD key data remains well below the slot limits of the homomorphic space where operations occur (typically 65,537 before reduction). The only minimal variations in size are due to random noise added to the ciphertext. In the traditional relay model, the first relay node does not execute any classical XOR operation on keys, whereas in the proposed architecture, this node performs an initial one-time pad encryption using XOR in the first homomorphic ciphertext, resulting in a small increase in its computational workload.

Furthermore, the fixed size of the payload, independent of key sizes or the number of hops, makes this method well-suited for large networks that consist of numerous relays. However, the proposed solution could face difficulties handling the final packet sizes for requests with more than one key per each, as supported by the ETSI-014 standard. An easy solution to avoid this situation could be to increase the number of requests by reducing the number of keys contained to a more suitable number or changing the default value of the maximum size limits in the web servers embedded in the relays.

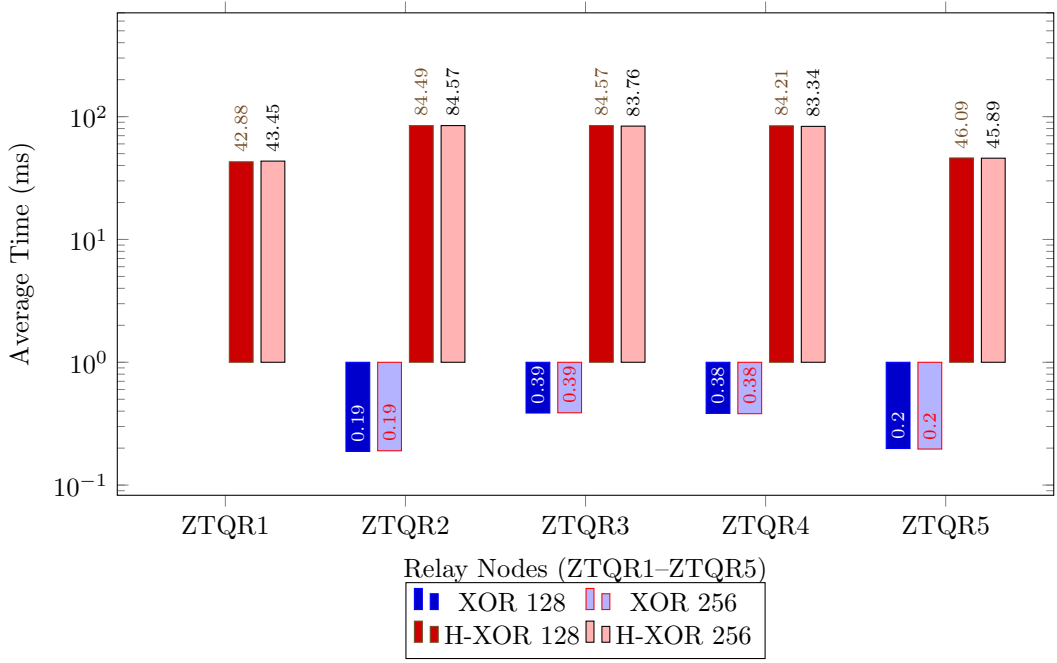
\begin{figure}[ht]
    \centering 
    \pgfplotstableread[col sep=comma]{table_data_times.csv}\finaltable
    
    \begin{tikzpicture}
        \begin{axis}[
          ybar,
          bar width=9pt,
          width=0.9\linewidth,
          height=8cm,
          symbolic x coords={ZTQR1, ZTQR2, ZTQR3, ZTQR4, ZTQR5},
          xtick={ZTQR1,ZTQR2,ZTQR3,ZTQR4,ZTQR5},
          xlabel={Relay Nodes (ZTQR1--ZTQR5)},
          ylabel={Average Time (ms)},
          ymode=log,
          log basis y=10,
          ymin=0,             
          ymax=700,           
          nodes near coords,
          point meta=rawy,
          every node near coord/.append style={
            font=\scriptsize,
            rotate=90,
            anchor=west
          },
          enlarge x limits=0.15,
          legend style={at={(0.5,-0.18)},anchor=north,legend columns=2},
        ]
          \addplot+[fill=blue!80!black, bar shift=-18pt, nodes near coords, every node near coord/.append style={text=white}]
            table[x=Node,y=XorOneHundredTwentyEight]{\finaltable};
          \addplot+[fill=blue!30!white, bar shift=-6pt]
            table[x=Node,y=XorTwoHundredFiftySix]{\finaltable};
          \addplot+[fill=red!80!black, bar shift=6pt]
            table[x=Node,y=HXorOneHundredTwentyEight]{\finaltable};
          \addplot+[fill=red!30!white, bar shift=18pt]
            table[x=Node,y=HXorTwoHundredFiftySix]{\finaltable};
        
          \legend{XOR 128, XOR 256, H-XOR 128, H-XOR 256}
        \end{axis}
    \end{tikzpicture}
    \caption{Operation Times Comparison: XOR vs. H-XOR}
    \label{fig:comparison_times}
\end{figure}

The processing time at each relay is the most notable change, as shown in Fig. \ref{fig:comparison_times}, increasing from approximately 0.2 ms for a standard XOR operation between two keys to nearly \textbf{45 ms}. Although this remains a manageable processing time, there are many opportunities for optimization. Since this implementation prioritizes functional demonstration and relies on technologies like Docker and Python, future versions could enhance performance by running the software directly on a Field Programable Gate Array (FPGA) chip, the host operating system or invoking OpenFHE from a compiled language such as C or Rust.

\begin{figure}[ht]
    \centering
    \begin{tikzpicture}
      \begin{axis}[
        boxplot/draw direction=y,   
        boxplot/every whisker/.style={solid}, 
        boxplot/every box/.style={solid}, 
        boxplot/every median/.style={solid}, 
        width=0.9\linewidth,
        height=9cm,
        xtick={1,2, 3,4,5,6, 7,8,9,10, 11,12,13,14, 15,16,17,18},
        xticklabels={
          H-XOR128,H-XOR256,  
          XOR128,XOR256,H-XOR128,H-XOR256,  
          XOR128,XOR256,H-XOR128,H-XOR256,  
          XOR128,XOR256,H-XOR128,H-XOR256,  
          XOR128,XOR256,H-XOR128,H-XOR256   
        },
        x tick label style={font=\small, rotate=90, anchor=east},
        ymin=0,
        ymajorgrids=true,
        xlabel={Relay Nodes (ZTQR1--ZTQR5)},
        ylabel={CPU Usage (\%)},
        after end axis/.code={
          \draw[dashed] (axis cs:2.5,0) -- (axis cs:2.5,3.7);
          \draw[dashed] (axis cs:6.5,0) -- (axis cs:6.5,3.7);
          \draw[dashed] (axis cs:10.5,0) -- (axis cs:10.5,3.7);
          \draw[dashed] (axis cs:14.5,0) -- (axis cs:14.5,3.7);
          \node at (axis cs:1.5,3.6) {ZTQR1};
          \node at (axis cs:4.5,3.6) {ZTQR2};
          \node at (axis cs:8.5,3.6) {ZTQR3};
          \node at (axis cs:12.5,3.6) {ZTQR4};
          \node at (axis cs:16.5,3.6) {ZTQR5};
        }
      ]
    
        \addplot+[
          boxplot,
          fill=red!20,
          draw=red!60!black
        ] table[y index=0]{cpu/tn-1-a2b2-performance-128bits-h-xor.csv};
    
        \addplot+[
          boxplot,
          fill=red!30,
          draw=red!60!black
        ] table[y index=0]{cpu/tn-1-a2b2-performance-256bits-h-xor.csv};
    
        \addplot+[
          boxplot,
          fill=blue!20,
          draw=blue!60!black
        ] table[y index=0]{cpu/tn-2-a2b1-performance-128bits-xor.csv};
    
        \addplot+[
          boxplot,
          fill=blue!30,
          draw=blue!60!black
        ] table[y index=0]{cpu/tn-2-a2b1-performance-256bits-xor.csv};
    
        \addplot+[
          boxplot,
          fill=red!20,
          draw=red!60!black
        ] table[y index=0]{cpu/tn-2-a2b1-performance-128bits-h-xor.csv};
    
        \addplot+[
          boxplot,
          fill=red!30,
          draw=red!60!black
        ] table[y index=0]{cpu/tn-2-a2b1-performance-256bits-h-xor.csv};
    
        \addplot+[
          boxplot,
          fill=blue!20,
          draw=blue!60!black
        ] table[y index=0]{cpu/tn-3-a1b3-performance-128bits-xor.csv};
    
        \addplot+[
          boxplot,
          fill=blue!30,
          draw=blue!60!black
        ] table[y index=0]{cpu/tn-3-a1b3-performance-256bits-xor.csv};
    
        \addplot+[
          boxplot,
          fill=red!20,
          draw=red!60!black
        ] table[y index=0]{cpu/tn-3-a1b3-performance-128bits-h-xor.csv};
    
        \addplot+[
          boxplot,
          fill=red!30,
          draw=red!60!black
        ] table[y index=0]{cpu/tn-3-a1b3-performance-256bits-h-xor.csv};
    
        \addplot+[
          boxplot,
          fill=blue!20,
          draw=blue!60!black
        ] table[y index=0]{cpu/tn-4-a3b4-performance-128bits-xor.csv};
    
        \addplot+[
          boxplot,
          fill=blue!30,
          draw=blue!60!black
        ] table[y index=0]{cpu/tn-4-a3b4-performance-256bits-xor.csv};
    
        \addplot+[
          boxplot,
          fill=red!20,
          draw=red!60!black
        ] table[y index=0]{cpu/tn-4-a3b4-performance-128bits-h-xor.csv};
    
        \addplot+[
          boxplot,
          fill=red!30,
          draw=red!60!black
        ] table[y index=0]{cpu/tn-4-a3b4-performance-256bits-h-xor.csv};
    
        \addplot+[
          boxplot,
          fill=blue!20,
          draw=blue!60!black
        ] table[y index=0]{cpu/tn-5-a4b4-performance-128bits-xor.csv};
    
        \addplot+[
          boxplot,
          fill=blue!30,
          draw=blue!60!black
        ] table[y index=0]{cpu/tn-5-a4b4-performance-256bits-xor.csv};
    
        \addplot+[
          boxplot,
          fill=red!20,
          draw=red!60!black
        ] table[y index=0]{cpu/tn-5-a4b4-performance-128bits-h-xor.csv};
    
        \addplot+[
          boxplot,
          fill=red!30,
          draw=red!60!black
        ] table[y index=0]{cpu/tn-5-a4b4-performance-256bits-h-xor.csv};
    
      \end{axis}
    \end{tikzpicture}
    \caption{CPU Usage Comparison}
    \label{fig:comparison_cpu}
\end{figure}
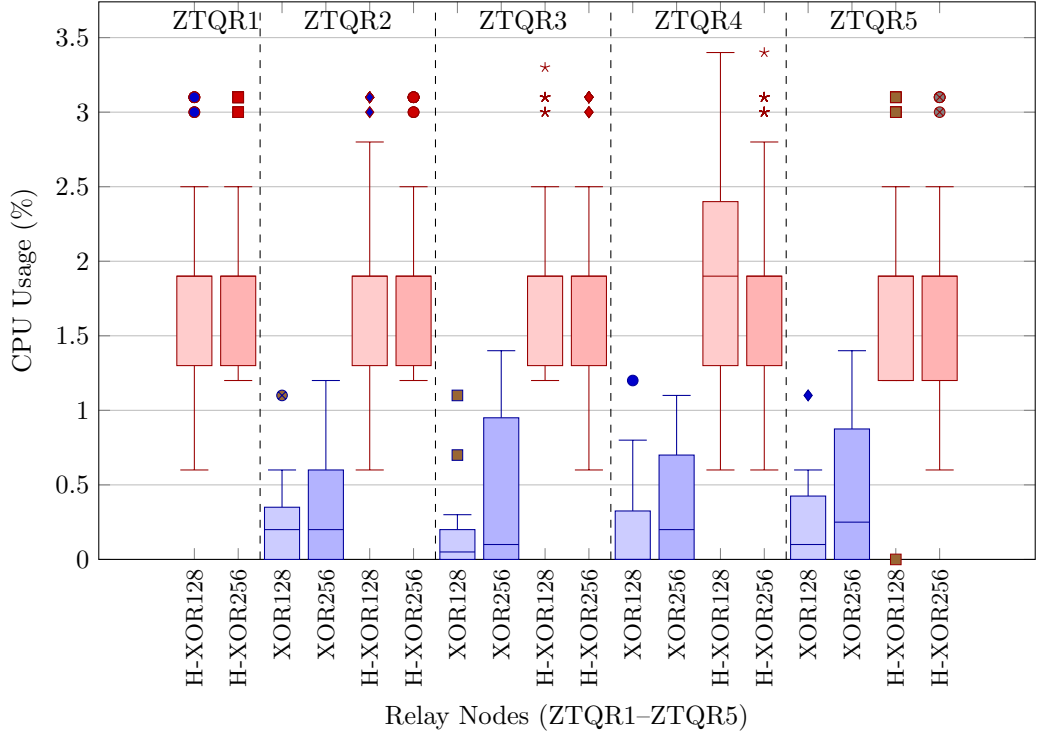

Regarding CPU usage, the tests indicate a noticeable overhead compared to traditional XOR, but the values are far from concerning. In our experiments, we observed an average CPU usage of \textbf{2\%}, measured from an isolated thread dedicated solely to key operations. The full range of CPU workload is shown in Fig. \ref{fig:comparison_cpu}. Since classic XOR is so lightweight that many measurements were close to 0\%, the H-XOR method introduces only a slight additional CPU load while offering a significant security advantage.

Another parameter to assess the scalability of the architecture is to prove that the noise growth related to the homomorphic additions applied to the encrypted keys is minimal enough to consider this solution scalable. For this purpose, some technical notation is needed. Firstly, the BFV scheme is based on a variant of LWE called Ring-LWE, because the plaintext and ciphertext spaces ($\mathcal{P}$ and $\mathcal{C}$, respectively) are based on mathematical rings. More precisely (Eq. \ref{eq:rings}):
\begin{equation}
    \mathcal{P}=\frac{\Z_p[x]}{x^n+1}\mbox{, }\mathcal{C}=\frac{\Z_q[x]}{x^n+1}\times\frac{\Z_q[x]}{x^n+1}.
    \label{eq:rings}
\end{equation}
Intuitively, elements in $\mathcal{P}$ are equivalent to vectors of length $n$ whose components are bounded modulo $p$, whereas elements in $\mathcal{C}$ are tuples $(c_1, c_2)$, where each $c_i$ is a vector of length $n$ whose components are bounded modulo $q$. Both $p$ and $q$ are integers such that $p <<q$. Following the homomorphic encryption standard \cite{hestandard}, for a 128-bit security, $q$ is a 220-bit prime number when $n=8192$ (the ring dimension by default for the BFV scheme in OpenFHE to achieve 128-bit security).

In order to show that the noise grows minimally any time two ciphertexts are added, it is important to note that a BFV public key lays in the same algebraic structure as the ciphertexts, $i.e.$, it is a tuple $(pk_1, pk_2)$ whose elements are randomly chosen vectors of length $n$ with components bounded modulo $q$. The encryption of a message $m(x)$ works as follows:
\begin{enumerate}
    \item Generate a random polynomial $u(x)$ of size $n$ with components chosen from the set $\{-1, 0, 1\}$. Generate two random polynomials $e_1(x)$ and $e_2(x)$ of size $n$ following a Gaussian distribution $N(0, 3.2)$ and bounded by 19 (refer to the HE standard in \cite{hestandard}). Compute $\Delta=\lfloor \frac{q}{p} \rfloor$, where $\lfloor\cdot\rfloor$ is the floor function.
    \item Compute $c_1(x)=pk_1(x)\cdot u(x)+e_1(x)+\Delta\cdot m(x)\mod q.$
    \item Compute $c_2(x)=pk_2(x)\cdot u(x)+e_2(x)\mod q.$
    \item The final ciphertext is $\textbf{c}=[c_1(x), c_2(x)]$
\end{enumerate}
Observe that all of $u(x)$, $e_1(x)$ and $e_2(x)$ are polynomials whose coefficients are always bounded by actually very low limits (compared to the 220-bit prime $q$). Given two ciphertexts $\textbf{c}$ and $\textbf{c}'$, adding them up is a relatively easy operation (Eq. \ref{eq:add}):
\begin{equation}
    \texttt{Add}(\textbf{c}, \textbf{c}')=[c_1(x)+c_1'(x), c_2(x)+c_2'(x)].
    \label{eq:add}
\end{equation}
Indeed, expanding the expression on the left (Eq. \ref{eq:expand}):
\begin{equation}
    c_1(x)+c_1'(x) = pk_1(x)\cdot [u(x)+u'(x)]+[e_1(x)+e_1'(x)]+\Delta\cdot [m(x)+m'(x)].
    \label{eq:expand}
\end{equation}
For $c_2(x)+c_2'(x)$ something similar happens (Eq. \ref{eq:expand2}):
\begin{equation}
    c_2(x)+c_2'(x) = pk_2(x)\cdot[u(x)+u'(x)]+[e_2(x)+e_2'(x)].
    \label{eq:expand2}
\end{equation}
In consequence, the addition of two ciphertexts results in another ciphertext $\textbf{c}''$. The only difference in $\textbf{c}''$ from a freshly encrypted ciphertext is that:
\begin{itemize}
    \item $u''(x)=u(x)+u'(x)$.
    \item $e_1''(x)=e_1(x)+e_1'(x)$.
    \item $e_2''(x)=e_2(x)+e_2'(x)$.
\end{itemize}
However, since all of the used elements are bounded by low limits compared to the 220-bit prime $q$, the added noise during a ciphertext addition can be thought as negligible. In other words, theoretically speaking, approximately $2^{220}$ additions can be made before having to worry about the key degradation.

In practice, though, a malicious ZTQR could try a noise injection attack, consisting of sending a very noisy ciphertext encapsulating their key, which cannot be decrypted at the end of the protocol. Recall from Section \ref{confidential} that multiplication is much more noisy than addition in FHE. Therefore, given $\mathbf{c}$ encrypting the keys and $\mathbf{c}_1$ encrypting $\mathbf{1}=(1, \dots, 1)$, one could evaluate $\mathbf{c}\cdot \mathbf{c}_1$ many times to obtain a ciphertext $\mathbf{c}_{\texttt{mal}}$ encrypting the key as expected but with rather much noise. Due to the large amount of products evaluated over $\mathbf{c}_{\texttt{mal}}$, the noise would be excessive for decryption. This is a well-known vulnerability of the proposed protocol, which can be addressed by the usage of zero-knowledge proofs of plaintext knowledge. Following the notation above, these proofs allow a prover to show a verifier that they know $m(x)$, $pk_1(x)$, $u(x)$ and $e_1(x)$ such that $c_1(x)=pk_1(x)\cdot u(x)+e_1(x)+\Delta\cdot m(x)\mod q$ for small $u(x)$ and $e_1(x)$, without revealing the original values. The proofs are directly related to the ciphertexts, so a malicious ZTQR cannot build such a proof for $\mathbf{c}_{\texttt{mal}}$. There already exist several zero-knowledge proofs of LWE plaintext knowledge in the literature that can be extended to the BFV scheme \cite{Lyubashevsky2017, delPino2019, Esgin2020, Lyubashevsky2022}.

In summary, the results presented in this section demonstrate the effectiveness of the proposed model, validating its potential in offering an alternative to the current trusted relay model without trust assumptions and supported by state-of-the-art technological components. The payload sizes and the noise induced in ciphertext demonstrate how this method can easily scale without problems in large QKD networks comprised of several quantum Links.

\section{Conclusions}
\label{sec:conclusions}
Future QKD networks equipped with quantum repeaters will probably not need the concept of classic relays. However, this evolution hinges on maturing quantum-repeater technology to a commercially viable level. Meanwhile, there remains a large installed base of QKD systems whose high cost and proprietary interfaces make them likely to persist in production environments for many years. As a result, overcoming the trusted relay model in QKDNs is one of the most relevant topics discussed in the community around this field. 

In this work, we propose an architecture that potentially solves the confidentiality concern in many of the QKD network deployments led by groups of nations like the EuroQCI project in the European Union \cite{EuroQCI}. Additionally, the presented architecture is based on existing standardisation efforts led by many industry partners and is potentially embeddable in QKD modules made by commercial vendors. 

Although several works are trying to solve this issue, as mentioned in section \ref{sec:related-work}. However, none, to our knowledge, has presented a solution with the same practical emphasis as the architecture proposed in this manuscript. Our proposal significantly improve the confidentiality of QKD networks without decreasing the security of the OTP key forwarding process, this contribution is specially interesting for large-scale QKD deployments such as EuroQCI and others. Homomorphic cryptography based on quantum-resistant algorithms has enough maturity level and is specially conceived to solve this kind of confidentiality challenge. Furthermore, the low complexity of the required operations to emulate an XOR function makes it a perfect solution to get all the benefits of this encryption schema without affecting the scalability of the solution. Another reason in favour of this approach is the non-affecting of the currently known quantum algorithms to the LWE family of cryptography, making it a formidable choice in order to keep the secrecy of the forwarded keys against "store now, decrypt later" attack patterns.

The presented protocol and architecture provide a relatively affordable way to face the most important confidentiality issues in the current QKDNs based on trusted relays. The core solution offers enough confidentiality to the relay operations and offers the possibility of improvement by the usage of specialised cryptographic hardware. Since the current ETSI-020 inter-KMS API is not present in many of the existing commercial equipment, the proposed method for confidentially exchanging keys between KMS can be used today as described, consuming keys from the existing interfaces and later migrated by vendors to the embedded QKD KMEs without making changes to the upcoming standard. However, this work does not address several open challenges that constitute potential directions for future research. These include the development of a quantum-resistant authentication scheme aligned with the latest post-quantum cryptography standards announced by NIST \cite{NIST2024}, the adaptation of this solution to key delivery protocols such as ETSI-004 \cite{ETSI004}, the optimisation of multiple-key management per request, and the examination of confidentiality challenges in state-of-the-art QKDN routing methods, all of which we intend to explore in future work.

\section*{Funding and acknowledgements}
This work was supported by the Basque Government through Plan complementario comunicación cuántica (EXP. 2022/01341) (A/20220551), the Spanish Ministry of Science and Innovation in the project EnablIng Native-AI Secure deterministic 6G networks for hyPer-connected envIRonmEnts (6G-INSPIRE) (PID2022-137329OB-C44), and in part by the project GN5-2 HORIZON-INFRA-2024-GEANT-01-SGA.

\section*{Disclosure}
The authors declare no conflicts of interest.

\bibliographystyle{ieeetr}
\bibliography{biblio}

\end{document}